\documentclass[9pt,twocolumn,twoside]{osajnl}

\journal{josab} 

\setboolean{shortarticle}{false}

\usepackage{color}

\usepackage[]{graphicx}
\usepackage{amsmath}
\usepackage{amssymb}
\usepackage{subfig}

\usepackage{epsfig}
\usepackage{wrapfig}
\usepackage{color}

\def\##1{{\bf #1}}
\def\=#1{\underline{\underline #1}}

\def\.{\mbox{ \tiny{$^\bullet$} }}

\def\les{\left[}
\def\ris{\right]}

\def\c#1{\cite{#1}}
\def\l#1{\label{#1}}

\def\epso{\epsilon_{0}}
\def\lambdao{\lambda_{ 0}}
\def\muo{\mu_{ 0}}
\def\ko{k_{ 0}}
\def\etao{\eta_{0}}

\def\eps{\varepsilon}
\def\epsa{\varepsilon_a}
\def\epsb{\varepsilon_b}

\def\pn{^{(n)}}
\def\pom{^{(m)}}
\def\bE{{\bf E}}
\def\bH{{\bf H}}
\def\br{{\bf r}}
\def\pp{^{(p)}}

\def\ux{\hat{\#u}_x}
\def\uy{\hat{\#u}_y}
\def\uz{\hat{\#u}_z}

\title{Rigorous formulation of surface plasmon-polariton-   waves propagation along the direction of periodicity of one-dimensional photonic crystal}

\author{Mehran Rasheed}
\author{Muhammad Faryad*}

\affil{Department of Physics, Lahore University of Management Sciences, Lahore 54792, Pakistan}

\affil[*]{muhammad.faryad@lums.edu.pk}

\begin{abstract}
A rigorous formulation of the canonical boundary-value problem is presented to find the surface plasmon-polariton waves guided by the interface of a one-dimensional photonic crystal  and metal along the direction of the periodicity. The problem is formulated  using the rigorous coupled-wave approach and a dispersion equation has been obtained. The Muller's method and Newton--Raphson method were used to obtain illustrative numerical results of the dispersion equation. The solutions were found to converge as the number of Floquet harmonics increased. This  formulation does not require  the computation of the photonic bandgaps and directly computes the surface-wave modes.  This   formulation could engender new applications of plasmonics exploiting the neglected interface along the direction of periodicity of the one dimensional photonic crystals. 
\end{abstract}

\setboolean{displaycopyright}{true}

\begin{document}

\maketitle

\section{Introduction}

The surface plasmon-polariton (SPP) waves are electromagnetic surface waves that propagate along the interface of a  dielectric material and a metal \cite{polo13} and find applications in optical sensing and imaging  \cite{mai07}. These applications emerge from the exploitation of the fact that the SPP waves have their field localized to the metal/dielectric interface. Electromagnetic surface waves between the interface of a homogeneous medium and one-dimensional photonic crystal (1DPC) were predicted in 1978 \cite{yeh78}. Since then, a lot of work has been done on these waves and their possible applications. The surface waves are often termed as Bloch surface waves when the partnering homogeneous medium is a dielectric material \cite{mea91} and are usually predicted using the surface band structures \cite{mea91,yab91}.   These surface waves are also called Tamm waves \cite{LPrev} after I. Tamm who discovered the electronic surface states of crystals \cite{ITamm}. When the partnering homogeneous medium is a metal, the surface waves are usually called just the SPP waves, but sometimes are called Tamm plasmon-polariton (TPP) in literature \cite{TPP1}. All these surface waves are localized to the interface plane that is perpendicular to the direction of periodicity of the 1DPC.

Surface waves guided  by the interfaces with the photonic crystals have various applications. For example, they can be used for highly directional emission \cite{kram04} and light could also be funneled efficiently through a photonic crystal \cite{mor04}.  High directional emission using two dimensional photonic crystals (2DPC) with a wide bandwidth have been studied \cite{li07} and directional beaming utilizing surface modes of 2DPC have also been investigated both theoretically and experimentally \cite{cag08}.

Surface waves on 1DPC has been utilized in designing biosensors \cite{kon07}, achieving negative refraction  \cite{bar08},  the  confinement of surface waves in one dimension \cite{wang17}, photonic crystal waveguides \cite{bagh18}, etc.\cite{sob18}.
Several theoretical studies have explored the effects of different aspects of the Bloch surface waves including the role of negative refraction  \cite{mou06}, volume surface waves \cite{hwa07}, the mode coupling and thickness of each layer of 1DPC \cite{arm03}, and TE surface waves on 1DPC \cite{mar06}.  Furthermore, several studies have shown that more than one SPP waves of the same frequency but of different polarization states and field profiles can exist on the interface of a metal and a 1DPC \cite{polo13,LPrev,far10,far12}.

However, most of the previous work on the surface waves on the interface of a homogeneous medium and the photonic crystal was either through photonic bandgaps \cite{mea91, yab91,  kram04, mor04, mou06, hwa07} or explored the properties of surface waves in 1DPC in the direction {\it perpendicular to the direction of the periodicity} of the 1DPC \cite{arm03, mar06, kon07, bar08, wang17, bagh18, sob18}. Here, we propose a  scheme that allows the computation of the properties of surface waves along the direction of periodicity of the 1DPC.  This scheme is extended from its use for the planar interface between a homogeneous medium and a 1DPC when the surface waves propagate perpendicular to the direction of the periodicity \cite{far10,LP2007,APA}. The formulation is developed using the rigorous coupled wave approach (RCWA) to solve Maxwell equations in the 1DPC. The RCWA uses the Fourier representation of the periodic permittivity and the fields to find the solution of the Maxwell equations and is very popular in finding the diffracted fields from gratings \cite{Glytsis,Chateau,L.L,M.G.Moh,Faryad-1}. This formulation does not require calculation of photonic bandgaps. For this purpose, a canonical boundary-value problem is set up that has a homogeneous medium in one half-space and the 1DPC in the other half-space.

The plan of the paper is as follow: The rigorous formulation of the   boundary-value problem is provided in Sec.~\ref{theory} and the numerical results are discussed in Sec.~\ref{nrd}. Concluding remarks are presented in Sec.~\ref{conc}.
For the formulation presented in this paper, an $\exp(-i\omega t)$ time-dependence is implicit, with $\omega$
denoting the angular frequency. The free-space wavenumber, the
free-space wavelength, and the intrinsic impedance of free space are denoted by $\ko=\omega\sqrt{\epso\muo}$,
$\lambdao=2\pi/\ko$, and
$\etao=\sqrt{\muo/\epso}$, respectively, with $\muo$ and $\epso$ being  the permeability and permittivity of
free space. Vectors are in boldface, 
column vectors are in boldface and enclosed within square brackets, and
matrixes are underlined twice and square-bracketed. The asterisk denotes the complex conjugate, the superscript
$T$ denotes the transpose,
and the Cartesian unit vectors are
identified as $\ux$, $\uy$, and $\uz$. 

\section{Formulation of dispersion equation}\label{theory}
\begin{figure}
\centering
\includegraphics[width=0.9\columnwidth]{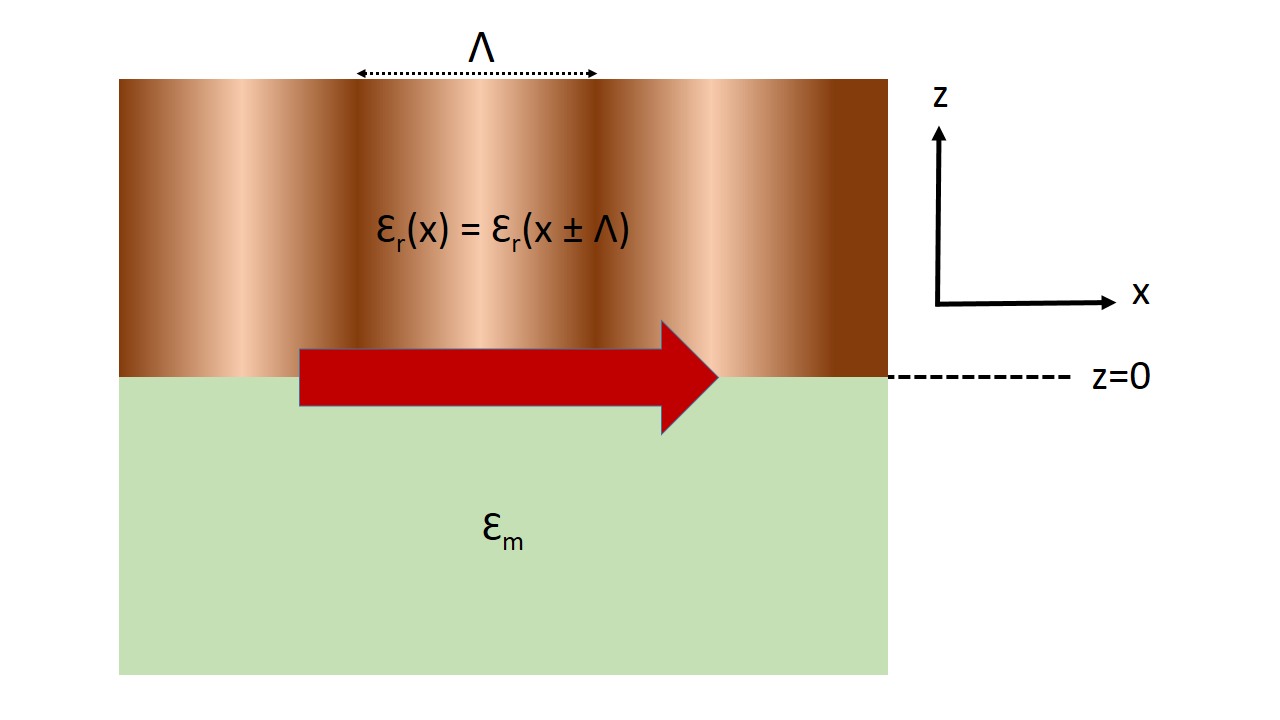}
\caption{{\bf Schematic for the canonical boundary-value problem:} Surface plasmon-polariton (SPP) waves (horizontal, red, thick arrow) are guided by a semi-infinite metal ($z\leq0$) with relative permittivity $\eps_m$ and a one-dimensional photonic crystal (1DPC)  occupying the half-space $z>0$ with relative permittivity $\eps_r(x) = \eps_r(x \pm \Lambda)$, where $\Lambda$ is the period of the 1DPC. } 
\label{geom} 
\end{figure}

Let us consider the boundary-value problem shown schematically in Fig. \ref{geom}. The half-space $z\leq0$ is occupied by a homogeneous metallic medium with permittivity $\eps_m$ and the region $z>0$ is occupied by a 1DPC of structural period $\Lambda$ along $x$ axis with the relative permittivity 
\begin{equation}
\eps_r(x)=\eps_r(x\pm\Lambda)\,,\quad z>0\,.
\end{equation}
The relative permittivity  can be expanded as a Fourier series with respect to $x$, viz.,
\begin{equation}
\eps_r(x)=\sum_{n\in \mathbb{Z}} \eps_r^{(n)}\exp (i n \kappa_x x)\,, \quad \mathbb{Z}\in\{0,\pm1,\pm2,...\}\,,
\label{perm}
\end{equation}
where  $\kappa_x=2\pi/\Lambda$,
\begin{equation}
\eps_r^{(0)}=
{1\over \Lambda}\int_0^\Lambda\eps_r(x) d x\,,
\end{equation}
and
\begin{equation}
\eps_r^{(n)}=
{1\over \Lambda}\int_0^\Lambda\eps_r(x) \exp(-in\kappa_xx)d x\,,\quad n\ne0\,.
\label{permn}
\end{equation}

In the half-space $z\leq 0$, let the SPP wave propagate in the $xz$ plane along  the $x$ axis. Since both mediums on either side of the interface plane $z=0$ are isotropic and we are considering the SPP-wave propagation along $x$ axis, both the $s$- and $p$-polarized waves decouple. Since only $p$-polarized SPP waves were found to exist in this geometry, we present below the formulation of the $p$-polarized SPP waves only.

 For the $p$-polarized SPP waves, the field phasors may be written in terms of Floquet harmonics as follows:
\begin{eqnarray}
\bE(\br)&=&{\sum_{n\in \mathbb{Z}}}a_p^{(n)}\left[-\frac{k_z^{(n)}}{\sqrt{\eps_{m}}\ko}\ux +{k_{x}^{(n)}\over\sqrt\eps_{m} \ko}\uz\right]\nonumber\\
&&\times\exp\left\{i\left[k_{x}^{(n)}x+k_z^{(n)}z\right]\right\}\,,\quad  z\leq0\,,\label{einc}\\[5pt]
{\etao}\bH(\br)&=& \sum_{n\in \mathbb{Z}}-a_p^{(n)}\sqrt{\eps_{m}}\uy
\exp\left\{i\left[k_{x}^{(n)}x+k_z^{(n)}z\right]\right\}\,,\quad  z\leq0\,,\nonumber\\\label{hinc}
\end{eqnarray}
where  $k_x\pn= q+n\kappa_x$ , $\kappa_x=2\pi/\Lambda$, 
and
\begin{equation}
k_z^{(n)}=\sqrt{
\eps_m\ko^2-\left[k_{x}^{(n)}\right]^2}\,\label{kz}
\end{equation}
is chosen such that ${\rm Im}\les k_z^{(n)}\ris<0$. Furthermore,
 $a_p^{(n)}$ are the unknown amplitudes and $q$ is unknown wavenumber of the SPP wave
 that have to be determined.

The field phasors may be written in the photonic crystal in terms of Floquet harmonics as 
\begin{equation}
\left.\begin{array}{l}
\bE(\br)=\displaystyle{\sum_{n\in \mathbb{Z}}} \,\les e_x\pn(z)\ux+e_z\pn(z)\uz\ris\exp\left[ik_x^{(n)}x\right]\\[5pt]
\bH(\br)=\displaystyle{\sum_{n\in \mathbb{Z}}}\, h_y\pn(z)\uy\exp\left[ik_x^{(n)}x\right]
\end{array}
\right\}\,.\label{field}
\end{equation}
Substitution of Eqs.~(\ref{perm}) and (\ref{field}) in the frequency-domain Maxwell curl postulates results in a system of three equations as follows:
\begin{eqnarray}
&&\frac{d}{dz}e_x\pn(z)-k_x\pn e_z\pn(z)=\ko{\etao} h_y\pn(z)\,,\label{max1}\\
&&\frac{d}{dz}h_y\pn(z)=\frac{\ko}{{\etao}}\sum_{m\in \mathbb{Z}}\epsilon_{r}^{(n-m)}e_x\pom(z)\,,\label{max5}\\
&&k_x\pn h_y\pn(z)=-\frac{\ko}{{\etao}}\sum_{m\in \mathbb{Z}}\eps_{r}^{(n-m)}e_z\pom(z)\,.\label{max6}
\end{eqnarray}

Equations~(\ref{max1})--(\ref{max6}) hold ${\forall}n\in \mathbb{Z}$.  We restrict $|n|\leq N_t$ and then define the column $(2N_t+1)$-vectors
 \begin{eqnarray}
 [\#x_\sigma(z)]&=&[x_\sigma^{(-N_t)}(z),~x_\sigma^{(-N_t)}(z),~...,\nonumber\\
 &&...,~x_\sigma^{(0)}(z),~...,~x_\sigma^{(N_t-1)}(z),~x_\sigma^{(N_t)}(z)]^T\,,
 \end{eqnarray}
 for $\#x\in\left\{\#e,\#h\right\}$ and $\sigma\in\left\{x,y,z\right\}$. Similarly, we define $(2N_t+1)\times(2N_t+1)$-matrixes
\begin{eqnarray}
[\=K_x]={\rm{diag}}[k_x\pn]\,,\qquad
[\=\eps_r]=\left[\eps_r^{(n-m)}\right]\,,
\end{eqnarray}
where ${\rm{diag}}[k_x\pn]$ is the diagonal matrix with $k_x\pn$ as the diagonal entries.

Equations (\ref{max6}) yield
\begin{eqnarray}
&&\left[\#e_z(z)\right] =-\frac{1}{\ko}\left[\=\eps_r\right]^{-1}\cdot\left[\=K_x\right]\cdot\left[\etao\#h_y(z)\right]
\end{eqnarray}
the use of which in Eqs.~(\ref{max1}) and (\ref{max5}) eliminates $e_z\pn\, \forall {n \in \mathbb{Z}}$, and gives the matrix equation
\begin{equation}
\frac{d}{dz}\left[\#f\pp(z)\right]=i[\=P\pp]\cdot\left[\#f\pp(z)\right]\,,\qquad z>0\,,\label{modep}
\end{equation}
where the column vector $[\#f\pp]$ with $2(2N_t+1)$ components is defined as 
\begin{equation}
\left[\#f\pp(z)\right]=\left[\left[\#e_x(z)\right]^T,~~\etao\left[\#h_y(z)\right]^T\right]^T
\end{equation}
and the $2(2N_t+1)\times2(2N_t+1)$-matrix $\left[\=P\pp\right]$ is given by
\begin{equation}
\left[\=P\pp\right]=\left[
\begin{array}{cccc}
\left[\=0\right] &\ko\left[\=I\right]-\frac{1}{\ko}\left[\=K_x\right]\cdot
\left[\=\eps_r\right]^{-1}\cdot\left[\=K_x\right]\\[5pt]
\ko\left[\=\eps_r\right]  & \left[\=0\right]
\end{array}
\right]\,,\label{Pmatp}
\end{equation}
where $\left[\=0\right]$ is the $(2N_t+1)\times(2N_t+1)$ null matrix and
$\left[\=I\right]$ is the  $(2N_t+1)\times(2N_t+1)$ identity matrix.

The solution of Eq. (\ref{modep}) can be written as
\begin{equation}
\left[\#f^{(p)}(z)\right]=\exp\left\{i[\=P^{(p)}]z\right\}\cdot\left[\#f^{(p)}(0)\right]\,,\qquad z>0\,.\label{modesolp}
\end{equation}
To write the field phasors of the surface waves in the photonic crystal, we need the field expressions that represent the decaying  fields away from the interface when $z \rightarrow\infty$. To do that, let us assume that $[\#t\pn]$ be the eigenvectors and $\alpha\pn$ be the corresponding eigenvalues of the matrix $2(2N_t+1)\times2(2N_t+1)$-matrix $\left[\=P^{(p)}\right]$ and are labeled such that the first $(2N_t+1)$ eigenvalues have Im $[\alpha\pn]>0$ so that the corresponding eigenvectors represent decaying fields as $z \rightarrow\infty$. The other half of the set of the eigenvalues represents the fields that grow as $z \rightarrow\infty$. Therefore, the field at the interface plane inside the 1DPC is
\begin{equation}
\left[\#f^{(p)}(0+)\right]=\left[[\#t^{(1)}],~[\#t^{(2)}],~...,~[\#t^{(2N_t+1)}]\right]\.[\#B^{(p)}]\,,\label{f0pp}
 \end{equation}
where 
\begin{equation}
[\#B^{(p)}] = \left[b_p^{(1)}, ~b_p^{(2)},~...,~b_p^{2N_t)},~b_p^{(2N_t+1)}\right]^T\,.
 \end{equation}

 Using Eqs. (\ref{einc}) and \ref{hinc}), we can get  the tangential components of the field phasors at $z=0-$ as
 \begin{equation}
 \left.
 \begin{array}{l}
 e_x\pn(0-)=-a_p^{(n)}k_z^{(n)}/\sqrt{\eps_{m}}\ko\\[4pt]
 \etao\,h_y\pn(0-)=-a_p^{(n)}\sqrt{\eps_{m}}
 \end{array}\right\}\,,\quad n\in\mathbf{Z}\,
 \end{equation}
that can be expressed in matrix form as
\begin{equation}
\left[\#f^{(p)}(0-)\right]=\left[
\begin{array}{c}
\left[\=Y^{(p)}_{e}\right]\\[5pt]
\left[\=Y^{(p)}_{h}\right]
 \end{array}\right]
 \. 
 \left[\#A^{(p)}\right]\label{f0mp}
 \end{equation}
where
\begin{eqnarray}
\left[\#A^{(p)}\right]=\big[a_p^{(-N_t)}, ~a_p^{(-N_t+1)},~...,~a_p^{(0)},~...,~a_p^{(N_t-1)},~a_p^{(N_t)}\big]^T\,,
\end{eqnarray}
and the $(2N_t+1)\times (2N_t+1)$-diagonal-matrixes $\left[\=Y_{e}^{(p)}\right]={\rm diag}\left[-k_z^{(n)}/\sqrt{\eps_{m}}\ko\right]$ and $\left[\=Y_{h}^{(p)}\right]=-\sqrt{\eps_{m}}\left[\=I\right]$.

 Implementing the standard boundary conditions, $[\#f^{(p)}(0+)]=[\#f^{(p)}(0-)]$, we get
 \begin{equation}
 [\=M^{(p)}]\. \left[
\begin{array}{c}
\left[\#B^{(p)}\right]\\[5pt]
\left[\#A^{(p)}\right] 
 \end{array}
 \right]\
= [\#0]\,,\label{M0p}
 \end{equation}
 where
\begin{equation}
 [\=M^{(p)}]=\left[[\#t^{(1)}],~[\#t^{(2)}],~...,~[\#t^{(2N_t+1)}~~
\begin{array}{c}
-\left[\=Y^{(p)}_e\right]\\[5pt]
-\left[\=Y^{(p)}_h\right] 
 \end{array}
 \right]\,.
 \end{equation}
 For non-trivial solutions, 
 \begin{equation}
{\rm det} [\=M^{(p)}]=0 \,,\label{disp}
 \end{equation}
 which is the dispersion equation of the SPP waves guided by the interface of a metal with the 1DPC along the direction of the periodicity.

\begin{figure}[h!t]
\centering
\includegraphics[width = 0.9\columnwidth]{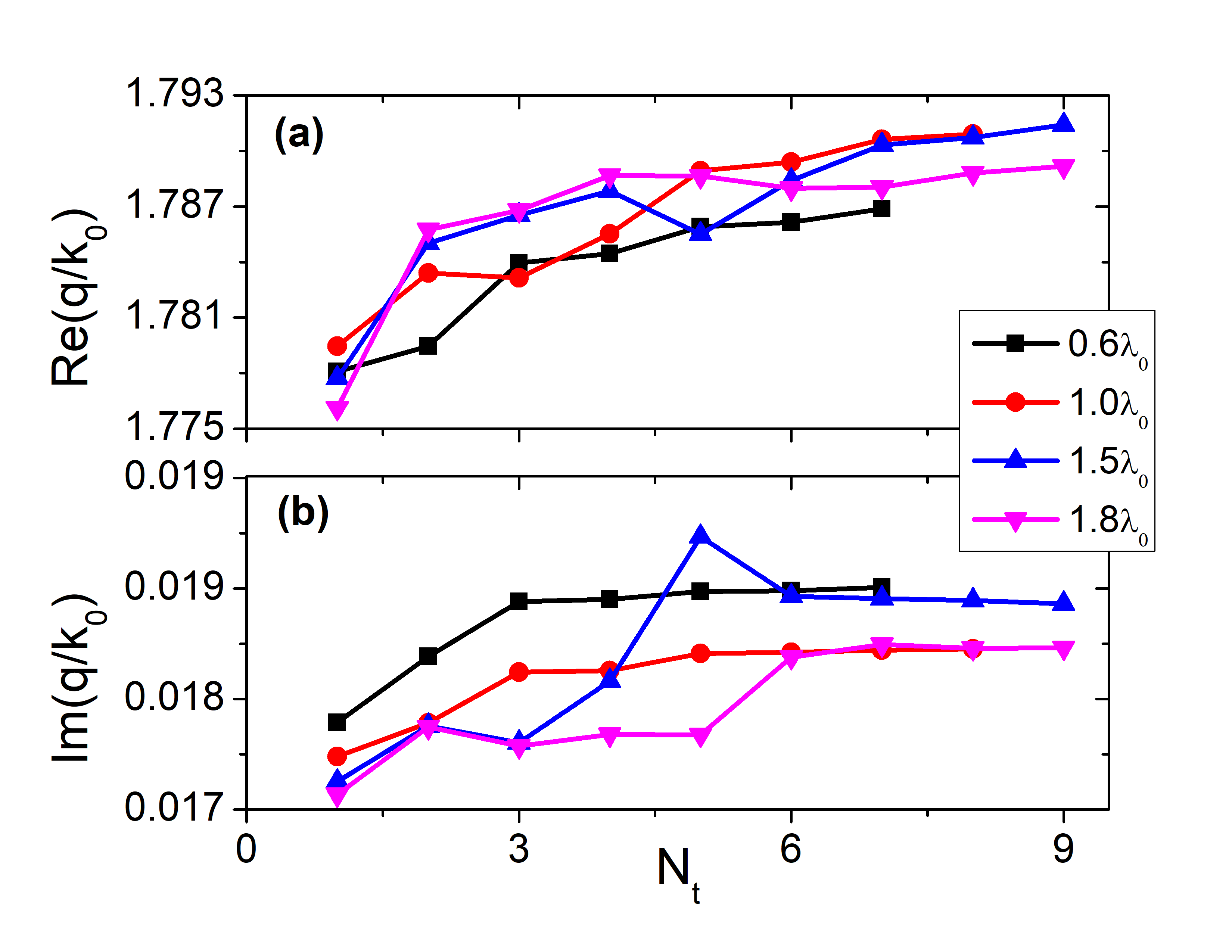}
\caption{ {\bf Solutions of dispersion equation (\ref{disp}):} The real and imaginary parts of the relative wavenumber as a function of $N_t$ when partnering metal is aluminum ($\eps_m = -54.7 + 21.9i$) and $\lambdao = 633$ nm,  $\epsa= (1.5)^2+10^{-6}i$ and $\epsb = (2)^2+10^{-6}i$ for different values of structural period $\Lambda$ of 1DPC.} 
\label{qs1} 
\end{figure}

\begin{figure}[h!t]
\centering
\includegraphics[width = 0.95\columnwidth]{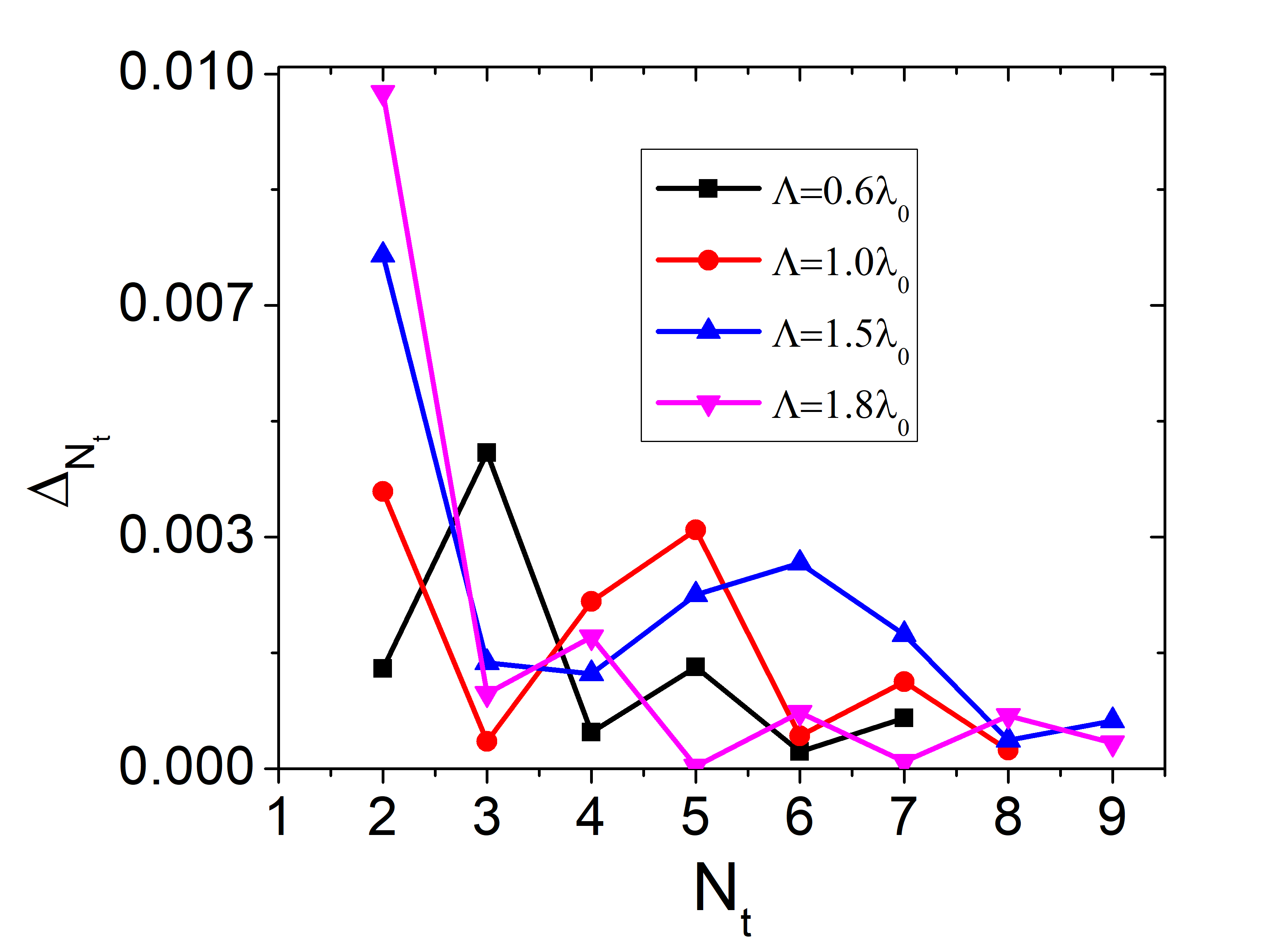}
\caption{ {\bf Convergence of solutions:} The absolute difference of the relative wavenumber (\ref{conv}) as a function of $N_t$ for the solutions presented in Fig. \ref{qs1}.} 
\label{qsdiff1} 
\end{figure}

\begin{figure}[h!t]
    \centering
        \includegraphics[width=0.95\columnwidth]{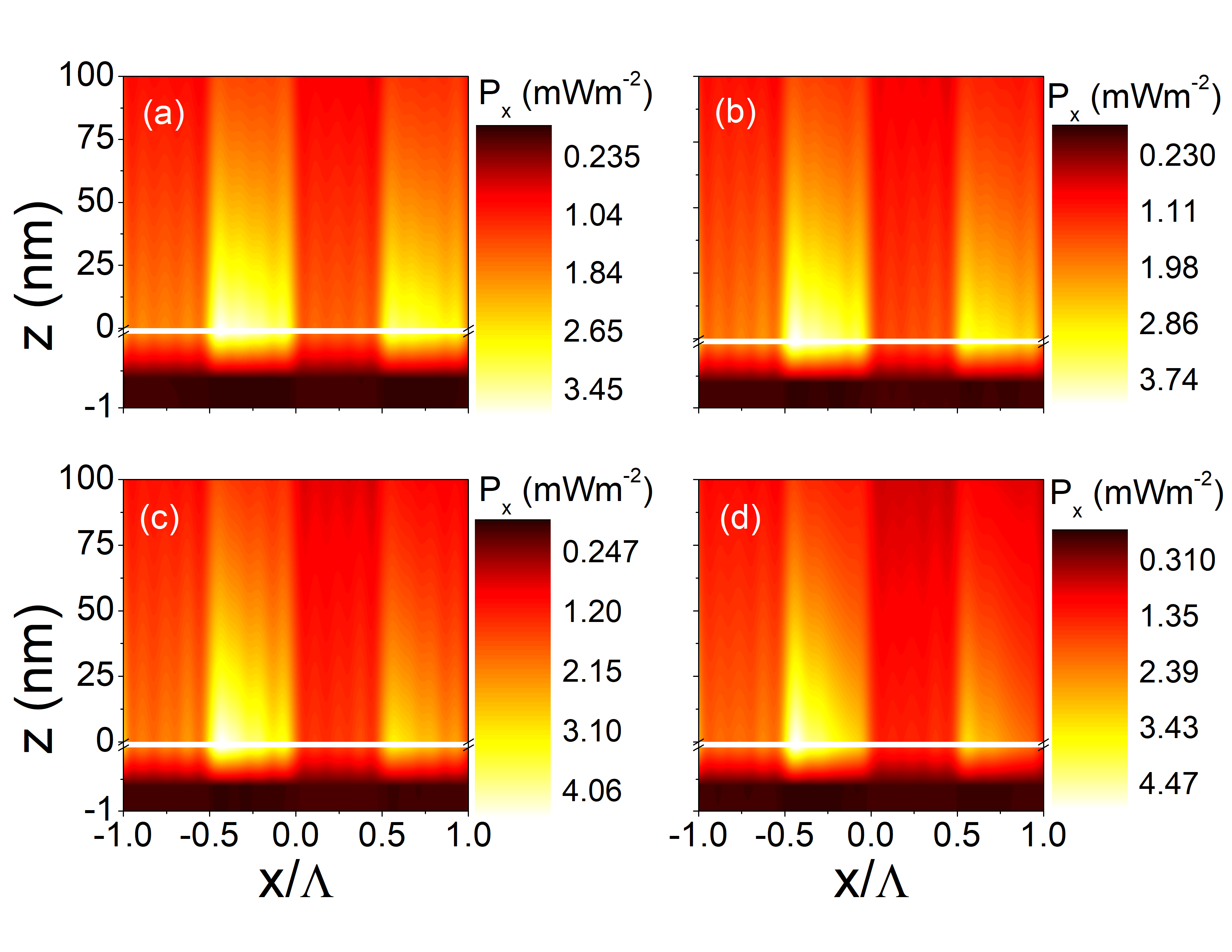}
        \caption{{\bf  Spatial profiles of power density:} Variation of the $x$-component of the time-averaged Poynting vector $P_{x}(x,z)$ as a function of $x$ and $z$ for the converged solutions of Fig. \ref{qs1} when $N_t = 7$ when  (a) $\Lambda = 0.6\lambdao$, (b) $\Lambda = 1.0\lambdao$, (c) $\Lambda = 1.5\lambdao$, and (d) $\Lambda = 1.8\lambdao$.} 
		\label{fp1} 
\end{figure}

\begin{figure}[h!t]
   \centering
        \subfloat{\includegraphics[width=0.45\columnwidth]{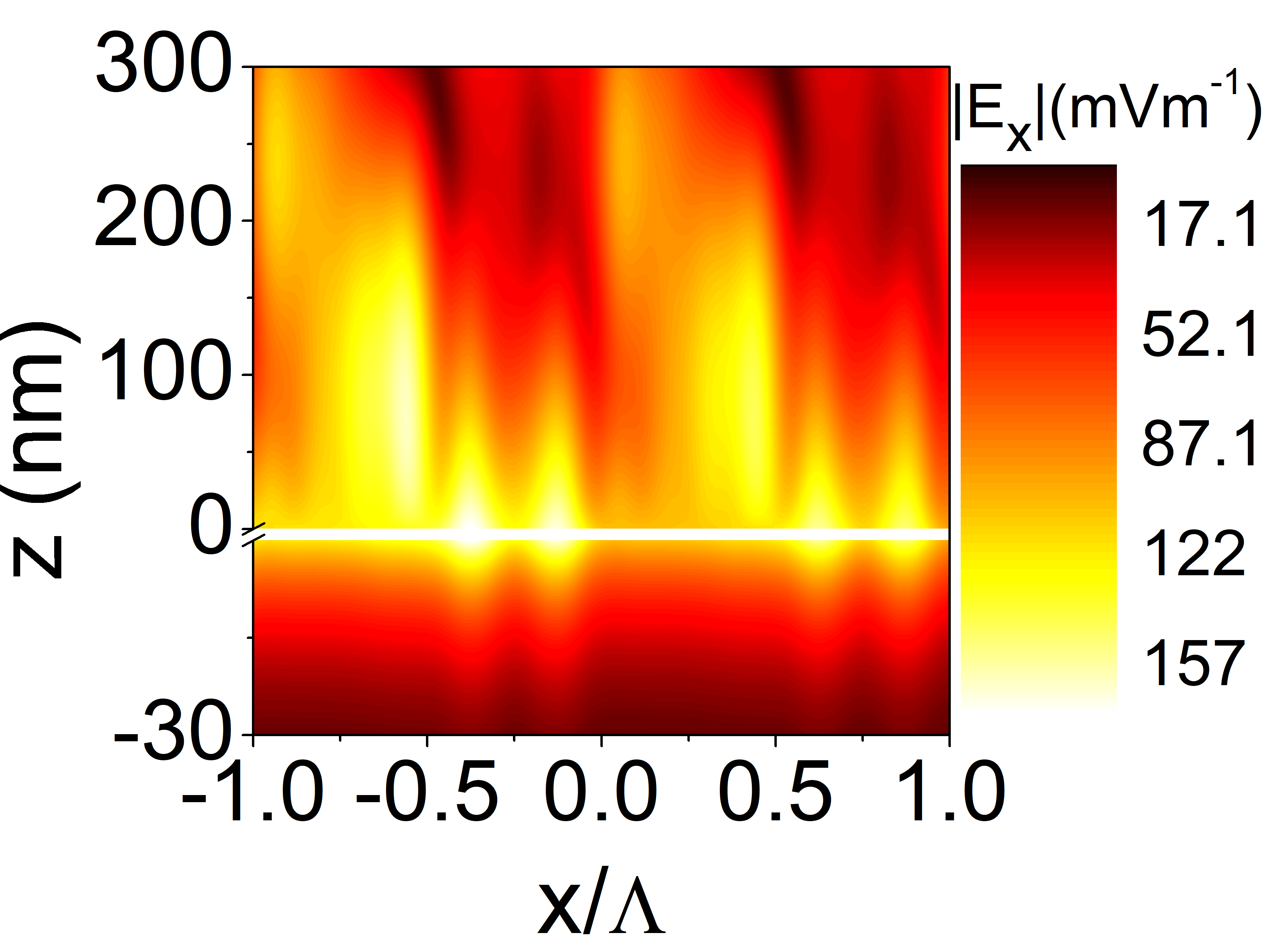}}
        \,\,
        \subfloat{\includegraphics[width=0.45\columnwidth]{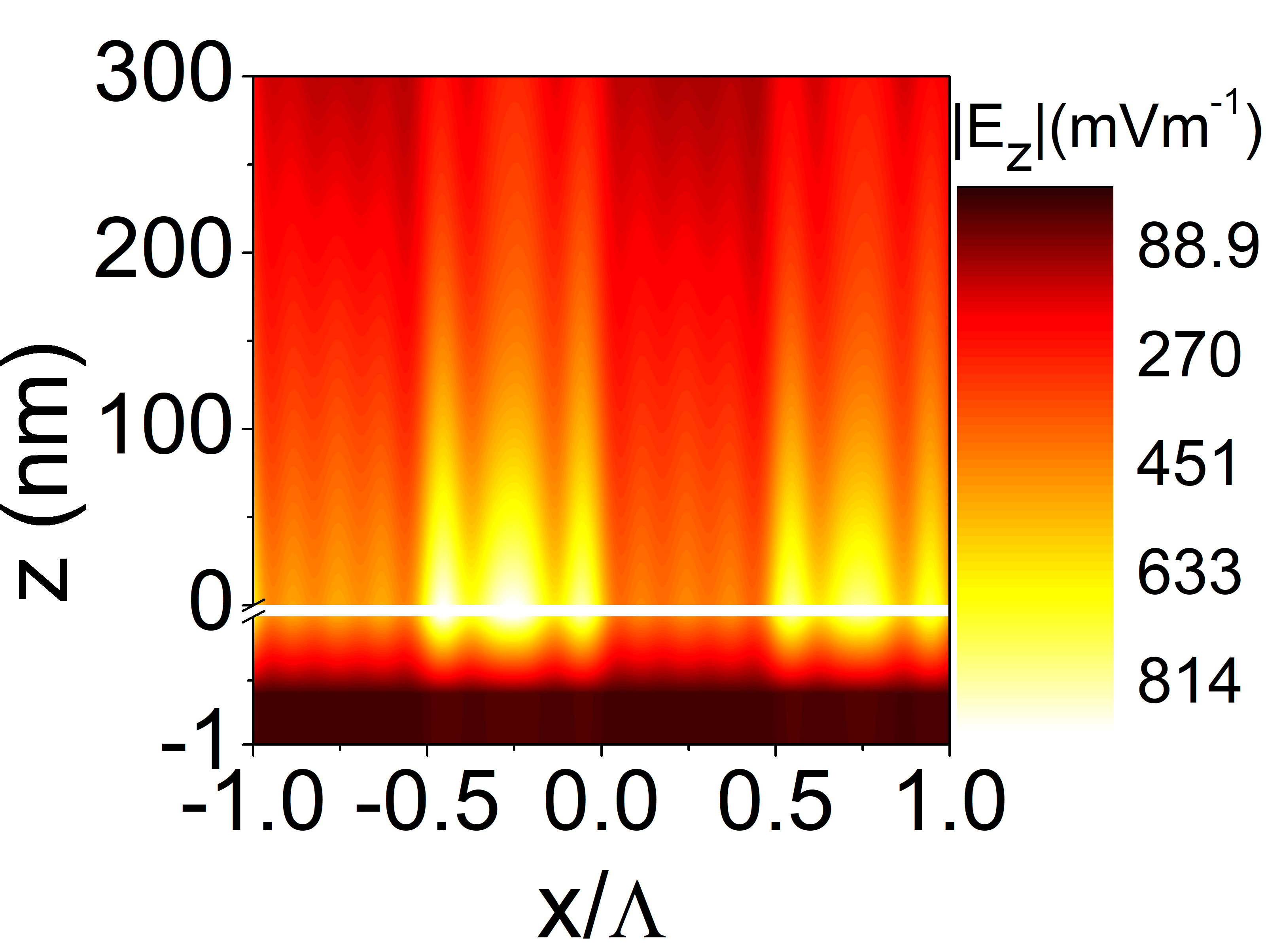}}
        \caption{{\bf  Spatial profiles of electric field:} Variation of (left)  the absolute value of the $x$-component of the electric field $|E_{x}|$ and (right) the absolute value of the $z$-component of the electric field $|E_{z}|$ as a function of $x$ and $z$ for the converged solutions of Fig. \ref{qs1} when $N_t = 7$ and $\Lambda = 1.0\lambdao$.}
		\label{ef1} 
\end{figure}

\section{Numerical Results and Discussion}\label{nrd}
For the representative numerical results, let us consider a 1DPC made of alternating layers of two dielectric mediums with relative permittivities $\eps_a$ and $\eps_b$ with both layers of equal thickness. Therefore,
\begin{equation}
\eps_r(x)=\left\{
\begin{array}{c}
\eps_a\,,~~~ 0<x<0.5\Lambda\,,\\
\eps_b\,, ~~~0.5\Lambda<x<\Lambda\,.
\end{array}
\right.
\end{equation}
The chosen 1DPC is not only the simplest to implement computationally but also the easiest to fabricate experimentally as alternating layers of dielectric materials are routinely fabricated using e-beam lithography and photo lithography. The Fourier coefficients of the permittivitiy are 
 \begin{equation}
\begin{split}
 \eps_r^{(0)} &= \frac{(\epsa + \epsb)}{2}\,,\qquad \eps_r^{(n)} = \frac{i(\epsb - \epsa)}{n\pi} \qquad \forall\, n = \{odd\} \,,\\
 &\eps_r^{(n)} = 0 \qquad \forall\, n = \{even\}\,.
\end{split}
 \end{equation}
For all the numerical results, the free-space wavelength was fixed $\lambdao = 633$ nm, the relative permittivities of the layers in the 1DPC were taken to be $\epsa= (n_a)^2+10^{-6}i$ and $\epsb = (n_b)^2+10^{-6}i$ with $n_a=1.5$ and $n_b=2$. The small imaginary part $10^{-6}$ is  included for the sake of numerical stability. The refractive index of $1.5$ and $2$ are chosen for the alternating layers since these are typical values of the dielectric materials in the visible range. For example, several silica based glasses have a refractive index of $1.5$ and $ZnO$ has a refractive index of $2$ at the chosen wavelength.

The dispersion equation (\ref{disp}) was solved using both the Muller's algorithm and the Newton--Raphson method \cite{Jaluria} and were found to be in excellent agreement. The initial guess solution for the wavenumber was taken to be
 \begin{equation}
 q/k_0 = \sqrt{ \frac{ \eps_m\eps_r^{(0)} }{ \eps_m+\eps_r^{(0)} } }\,,\label{shde}
 \end{equation}
and the solution was found when the real and imaginary parts of ${\rm det} [\=M^{(p)}]$ were found to be smaller than $10^{-4}$. The convergence of the solution was also investigated as a function of the number of terms $2N_t+1$. Two types of metallic partnering mediums were taken for illustrative results, as discussed below. Let us note that the Muller's algorithm gave all the results that are presented here. Newton--Raphson method also gave the same results whenever it converged, which was usually the case for smaller value of $N_t$.

\begin{figure}[h!t]
\centering
\includegraphics[width=0.95\columnwidth]{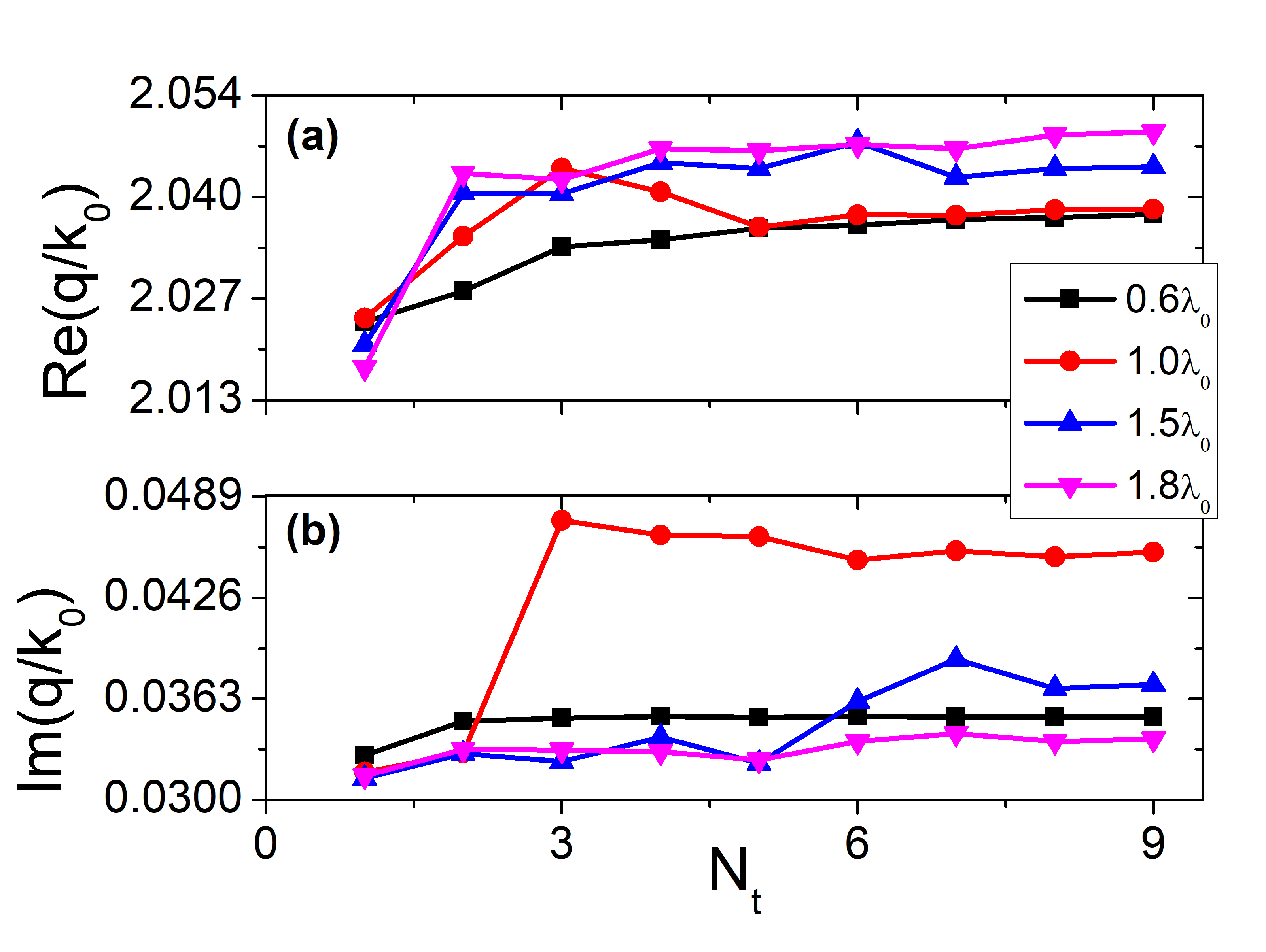}
\caption{Same as Fig. \ref{qs1} except that the partnering metal is gold ($\eps_m = -11.8 + 1.3i$).} 
\label{qs2} 
\end{figure}

\begin{figure}[h!t]
\centering
\includegraphics[width=0.9\columnwidth]{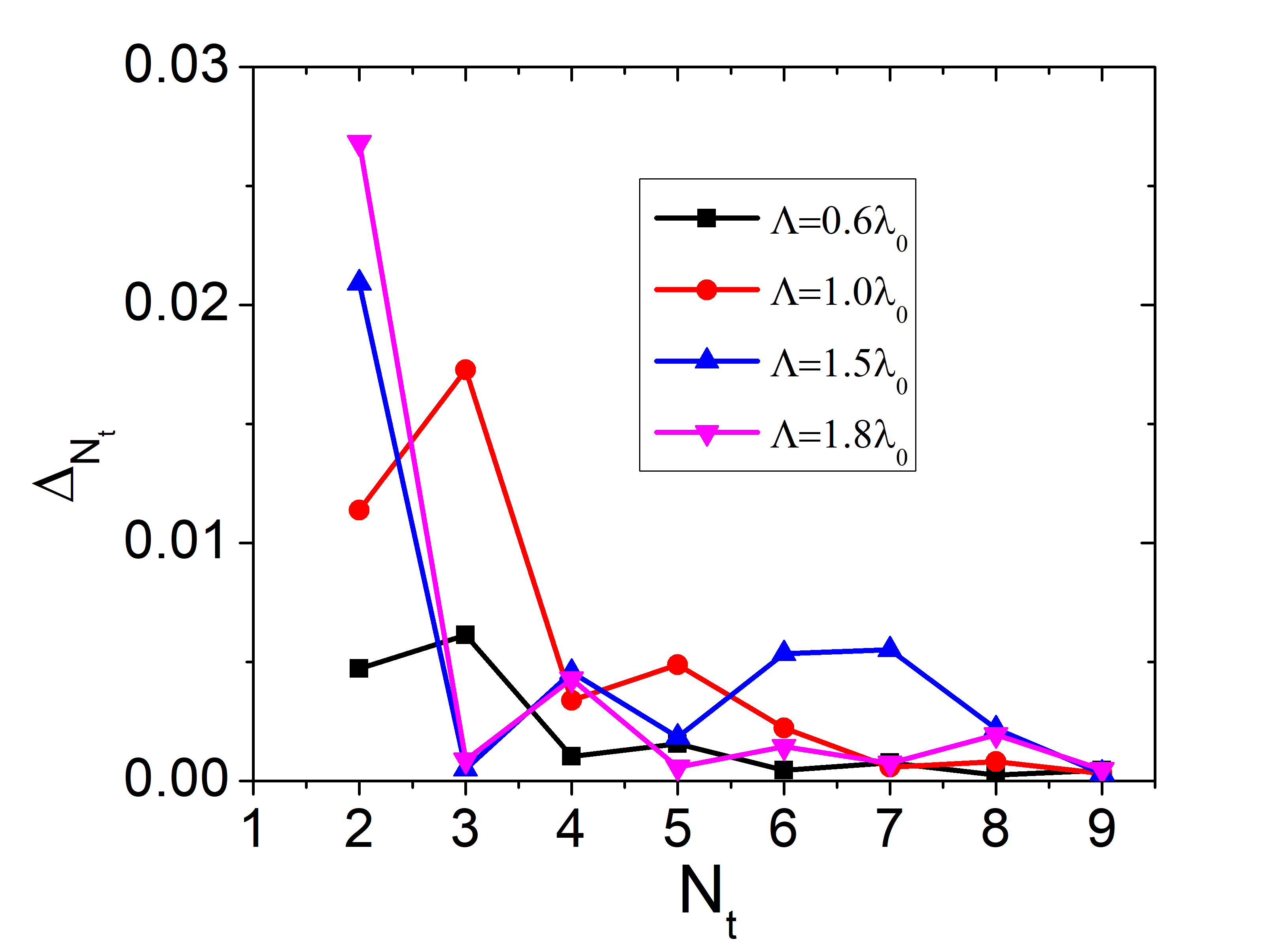}
\caption{Same as Fig. \ref{qsdiff1} except that the partnering metal is gold ($\eps_m = -11.8 + 1.3i$) and the convergence of the solutions of Fig. \ref{qs2} is presented.} 
\label{qsdiff2} 
\end{figure}
\begin{figure}[h!t]
    \centering
        \includegraphics[width=0.95\columnwidth]{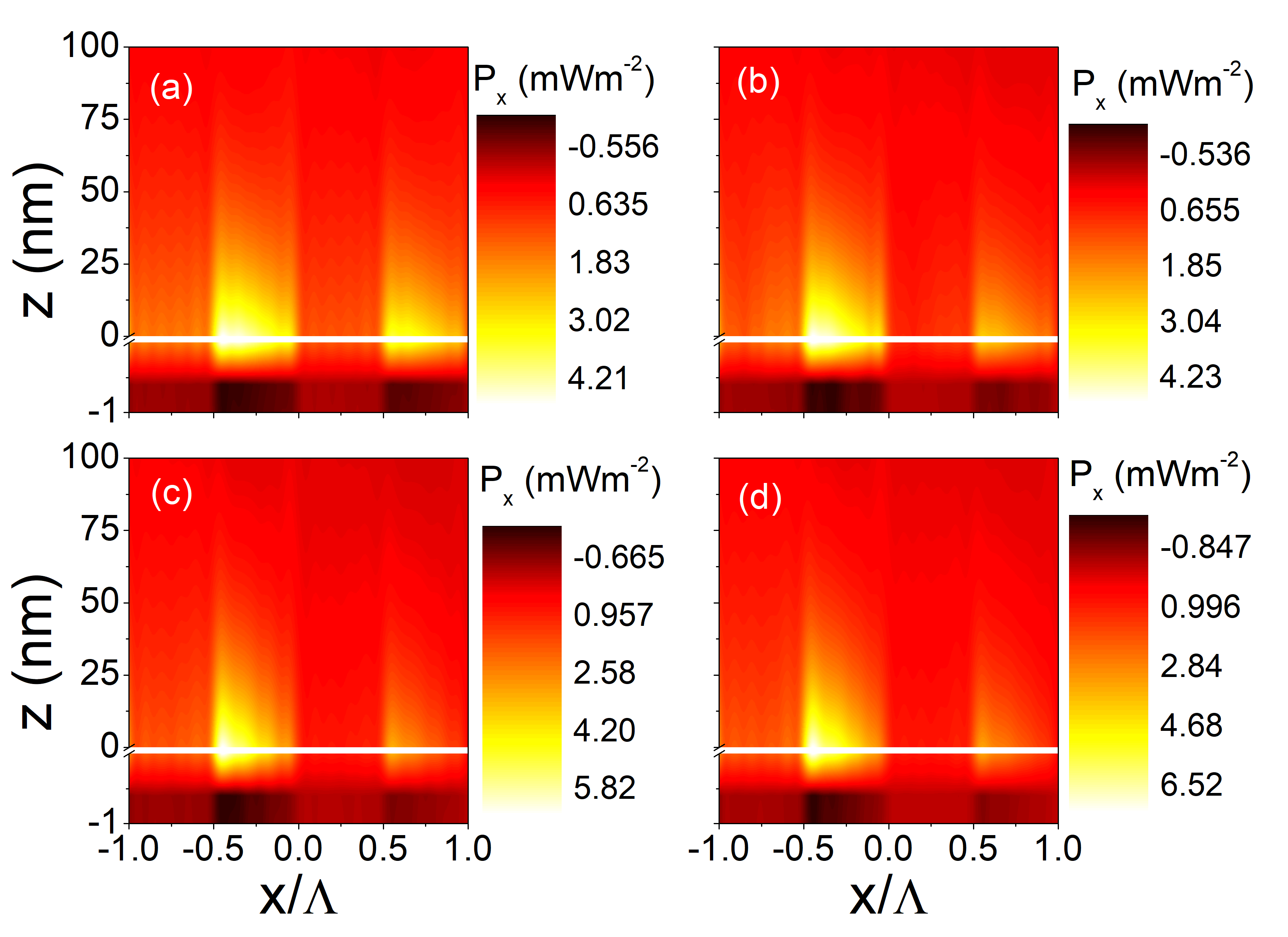}
        \caption{ Same as Fig. \ref{fp1} except that the partnering metal is gold and $N_t=9$.} 
		\label{fp2} 
\end{figure}

 \subsection{Aluminum/1DPC interface}
 Figure \ref{qs1} shows the solution of the dispersion equation (\ref{disp}) for different periods of $\Lambda$ of the 1DPC as a function of $N_t$ when the partnering metal is assumed to be aluminum with $\eps_m = -54.7+21.9i$~\cite{ARef}.  It can be seen that the solutions converge as $N_t$ increases. Let us also note that the both numerical algorithms were found to be unstable beyond a certain value of $N_t$ {\it after} the solution has converged. Usually, this value of $N_t$ was smaller for smaller $\Lambda$ as is also evident from Fig. \ref{qs1} because the solution for $\Lambda=0.6\lambdao$ were found until $N_t=7$ and the solutions for $\Lambda=\lambdao$ were found until $N_t=8$. However, the solutions for these values have converged quite considerably. An important feature of the SPP waves is that $n_a<\sqrt{\eps_r^{(0)}}<{\rm Re}\left(q/\ko\right)<n_b$. Therefore, the ``wavelength'' of the SPP wave is larger than that of the propagating waves in the medium with the permittivity equal to the average permittivity $\eps_r^{(0)}$
of the 1DPC. In order to emphasize the fact that the solutions of the dispersion equation (\ref{disp}) indeed converge, we plotted the absolute difference of the relative wavenumbers 
\begin{equation}
\Delta_{N_t} =\mid\!(q/k_0)_{N_t}-(q/k_0)_{N_t-1}\!\mid\,.\label{conv}
\end{equation}
 in Fig. \ref{qsdiff1}. The figure shows that the absolute relative difference is indeed decreasing with an increase in $N_t$.

To ascertain if the solutions of the dispersion equation indeed represent SPP waves, we computed the spatial field profiles using Eqs. (\ref{einc}), (\ref{hinc}) and (\ref{modesolp}) after finding the unknown coefficients from Eq. (\ref{M0p}). From the fields, the  $x$-component of the time-averaged Poynting vector
\begin{equation}
P_x(x,z)=0.5{\rm Re}\left[\#E(x,z)\times\#H^\ast(x,z)\right]\cdot\ux\,
\end{equation}
as a function of $x$ and $z$ since this component represent the power density flowing along the direction of propagation of the SPP waves and is presented in Fig. \ref{fp1} for four values of $\Lambda$ for the converged solutions of Fig. \ref{qs1} when $N_t=7$. The figure shows strong localization to the interface $z=0$ and the decay of fields as $x$ increases since $q$ is complex, but the decay is according to Floquet--Layponov theorem \cite{YS75}, i.e., the envelop of a periodic variation is decaying. Furthermore, the figure shows that the fields are much stronger in $b$ layer (with refractive index of $2$) than in $a$ layer (with refractive index $1.5$) though the field decays away from the interface $z=0$ in both the layers. Let us also note that the field is very strongly localized in the metallic side of the interface as compare to the common case of SPP wave at the interface homogeneous dielectric and metal. 

Figure \ref{ef1} shows the variation of the absolute value of the $x$-component of the electric field $|E_{x}(x,z)|$ and the absolute value of the $z$-component of the electric field $|E_{z}(x,z)|$ as a function of $x$ and $z$. It can be seen that $E_z$ decays more quickly as compared to $E_x$, but again the decay is according to Floquet--Layponov theorem \cite{YS75}, i.e., the envelop of a periodic variation is decaying. The comparison of the two panels show that the the $E_x$ is stronger than $E_z$.

\subsection{Gold/1DPC interface}
Figure~\ref{qs2} shows the solution of the dispersion equation (\ref{disp}) for the gold/1DPC interface using the same photonic crystal. The relative permittivity for the gold was taken to be $-11.8 + 1.3i$ \cite{GRef}. The figure confirms that the solutions converge as $N_t$ increases. However, in this case $n_a<\sqrt{\eps_r^{(0)}}<n_b<{\rm Re}\left(q/\ko\right)$ in contrast to the case of aluminum. Also, the SPP waves are much more lossy as compared to the case of aluminum as can be seen from a comparison of Figs. \ref{qs1}(b) and \ref{qs2}(b) since the ${\rm Im}\left(q/\ko\right)$ for gold is about five times larger than that for the aluminum. The absolute of the differences of the relative wavenumbers (\ref{conv}) in Fig. \ref{qsdiff2} shows that the solutions of the dispersion equation (\ref{disp}) converge with an increase in $N_t$.

The spatial profiles of the power density of the SPP waves are shown in Fig. \ref{fp2} for four values of the period $\Lambda$ when $N_t=9$. The figure shows that the SPP waves are very strongly localized to the $z=0$ interface on both sides. Also, the decay of the power density with $x$ is also much higher as compared to the case of aluminum because of higher values of ${\rm Im}\left(q/\ko\right)$ for gold. As for the case of aluminum, the fields are much stronger in $b$ layer than in $a$ layer. 

\section{Concluding Remarks}\label{conc}
A rigorous formulation for computing the wavenumbers and the spatial field profiles of the surface plasmon-polariton (SPP) waves guided along the direction of periodicity of a one-dimensional photonic crystal (1DPC) was developed and the representative numerical results were computed. The formulation does not require the photonic bandgap structure and directly solves the eigenvalue problem. The numerical results show that the SPP waves are highly localized to the interface of a metal and the 1DPC. It is hoped that this formulation will open up new avenues of research to exploit this long-neglected interface since most of the research has been focused on the other interface where the surface wave propagates in an interface perpendicular to the direction of periodicity. Let us note that this scheme can be extended to find canonical solutions of possible SPP waves guided by one-dimensional gratings combining the technique presented in this paper with the techniques of surface waves guided by a slab of one material sandwitched between two other materials \cite{MLjouk,MLpra1,meh17}.

\section*{Funding Information}

This work is partially supported by the Higher Education Commission of Pakistan (HEC) grant NRPU $5905$.

\end{document}